\documentclass[twocolumn,superscriptaddress,showpacs,amsmath,amssymb,pre]{revtex4}

\usepackage{bm}
\usepackage{times}

\begin{document}

\newcommand{\beq}{\begin{equation}}
\newcommand{\eeq}{\end{equation}}
\newcommand{\bea}{\begin{eqnarray}}
\newcommand{\eea}{\end{eqnarray}}
\newcommand{\req}[1]{Eq.\ (\ref{#1})}
\newcommand{\gcc}{{\rm~g\,cm}^{-3}}

\title{Comment on ``Equation of state of dense and magnetized fermion system''}
\author{A. Y. Potekhin}
\affiliation{Ioffe Physical-Technical Institute
of the Russian Academy of Sciences,
Politekhnicheskaya 26, 194021 St.~Petersburg, Russia}
\affiliation{CRAL (UMR CNRS No. 5574), 
Ecole Normale Sup\'{e}rieure de Lyon,
69364 Lyon Cedex 07, France}

\author{D. G. Yakovlev}
\affiliation{Ioffe Physical-Technical Institute of the Russian
Academy of Sciences, Politekhnicheskaya 26, 194021 St.~Petersburg,
Russia}
\affiliation{Saint-Petersburg State Polytechnical University,
Politekhnicheskaya 29, 195251 St.\ Petersburg, Russia}

\date{\today}
\begin{abstract}
Contrary to what is claimed by Ferrer et al.\ [Phys.\ Rev. C
\textbf{82}, 065802 (2010)], the magnetic field of a neutron star
cannot exceed $10^{19}$~G and the thermodynamic pressure of dense
magnetized fermion gas is isotropic.
\end{abstract}

\pacs{21.65.Mn, 26.60.Kp, 97.60.Jd}

\maketitle


The authors of recent paper \cite{ferreretal10} construct
thermodynamics of charged fermions in strong magnetic field
$\bm{B}$ where Landau quantization of orbital motion is
important and thermodynamic quantities depend on $\bm{B}$.
The subject attracts considerable attention, with the most
important applications to neutron stars possessing strong
magnetic fields. The authors conclude that (i) the magnetic
field in a neutron star can exceed $10^{19}$~G and (ii) the
gas of particles in a quantizing magnetic field has
anisotropic pressure. We point out that both statements are
inaccurate.

\section{Maximum field strength}

The well known estimate based on the virial theorem
\cite{ChandraFermi} gives the maximum magnetic field in a
neutron star $B_\mathrm{max}\sim10^{18}$~G
\cite{LaiShapiro}. The authors of Ref.~\cite{ferreretal10}
claim that this estimate can be relaxed. As an alternative,
they propose arbitrary simplistic parametrizations of mass
density $\rho$ and field strength $B$ as functions of the
radial coordinate $r$ within the star, treating the
parameters of these functions as ``totally arbitrary.'' For
certain values of these parameters they obtain
$B_\mathrm{max}>10^{19}$~G.

However, the density and field distributions are not
arbitrary, but must satisfy stability equations for a
magnetized star with a realistic equation of state. Detailed
self-consistent numerical simulations (for example,
\cite{Bocquet,KK08}) show that, depending on the adopted
equation of state in the stellar core, $B_\mathrm{max}$
takes values $(0.3-3.0)\times10^{18}$~G, in disagreement
with Ref.~\cite{ferreretal10} but in good agreement with
Ref.~\cite{LaiShapiro}. A large variety of equations of
state were explored in numerical simulations \cite{Bocquet}.
The obtained $\rho$ and $B$ distributions are different from
artificial distributions of Ref.~\cite{ferreretal10},
leading to different values of $B_\mathrm{max}$.

\section{Isotropy of pressure}

The consideration of the pressure in
Ref.~\cite{ferreretal10} is based on the articles by Canuto
and Chiu \cite{cc68} who showed that kinetic pressures
$p_\|^\mathrm{kin}$ and $p_{\perp}^\mathrm{kin}$ of charged
particles along and across $\bm{B}$, calculated as ensemble
averages of respective currents of kinetic momenta, are
different. The authors of Ref.~\cite{ferreretal10} repeat
the consideration \cite{cc68} using a more general formalism
and arrive at the same conclusions. According to
Refs.~\cite{cc68,ferreretal10}, the total anisotropic
pressure is the sum of the magnetic pressure related to the
Maxwell stress tensor, and the kinetic pressure. The
longitudinal and transverse
kinetic pressures are
$p_\perp^\mathrm{kin} = -\Omega-M B$ and $p_\|^\mathrm{kin}
= -\Omega$,
where $\Omega$ is the grand
canonical potential per unit volume and $\bm{M}$ is the
magnetization (directed along $\bm{B}$ in the
quasistationary approximation adopted in these studies).

However, the deficiency of the approach of Ref.~\cite{cc68} has been
pointed out long ago by Blandford and Hernquist \cite{bh82}. It is
well known that the total microscopic electric current density
$\bm{j}$ is composed of the free (or conduction) current term
$\bm{j}_f$ and bound current term $\bm{j}_b$ due to magnetization
(dynamical polarization contribution to $\bm{j}_b$ in the
quasi-stationary approximation is negligible). The magnetization
current density equals (in Gaussian units) $\bm{j}_b=c\,\nabla \times
{\bm{M}}$; in case of boundaries, this volume current should be
supplemented by the surface current $c\bm{M}\times\bm{B}/B$ (see,
e.g., Ref.~\cite{Griffith}). The total \textit{thermodynamic}
pressure $P$ in a magnetized plasma is the sum of the kinetic
pressure and an additional contribution due to the Lorentz force
density related to the magnetization currents. If we compress a
plasma across $\bm{B}$, then the magnetization current density
induces an additional contribution $MB$ to the force density. As a
result, the transverse component of the total (thermodynamic) plasma
pressure equals $p_\perp^\mathrm{kin}+MB=p_\|^\mathrm{kin}$, so that
the total plasma pressure $P=-\Omega$ is isotropic.

In spite of simplicity of the above arguments, they are
sometimes ignored in the literature, like in
Ref.~\cite{ferreretal10}. Therefore, in order to make them
still more transparent, let us illustrate the pressure
isotropy with two graphic examples.

As the
simplest example, consider a plasma contained in a finite cylinder in
vacuum with a uniform external $\bm{B}$-field along the cylinder
axis. At equilibrium
in the absence of external
forces, the sum of the force densities exerted on the side wall of
the cylinder by transfer of kinetic momenta of plasma particles and
by the surface magnetization current equals
$p_\perp^\mathrm{kin}+MB=-\Omega$. It is the same as the force
density $p_\|^\mathrm{kin}=-\Omega$ exerted on the head wall. Hence
the plasma pressure, which can be
determined in this experiment by measuring forces on the cylinder
walls, is isotropic.

As another example, more relevant to astrophysics, consider
a volume element in a magnetized star. Let the element be
sufficiently small and distributions of $\bm{B}$,
temperature $T$, and gravitational acceleration
$\bm{g}$ be sufficiently smooth, so that we can assume constant
$\bm{B}$, $T$, and $\bm{g}$ within this volume. Let the $z$ axis be
directed along $\bm{g}$. Then $\rho$ and $\Omega(\rho,B,T)$ depend on
$z$, resulting in $z$-dependent magnetization
$\bm{M}=-\partial\Omega(\rho,\bm{B},T)/\partial \bm{B}$. Hydrostatic
balance implies the density of gravitational force, $\rho \bm{g}$, be
balanced by the density of forces created by plasma particles
(gradient of kinetic pressure and Lorentz force due to plasma
magnetization).

Now let us compare two limiting cases. If $\bm{B}$ is
parallel to $\bm{g}$, the $z$-component of Lorentz force is
absent, and we
get the standard equation of
hydrostatic equilibrium $\rho
g=dp_\|^\mathrm{kin}/dz=dP/dz=-d\Omega/dz$.

If $\bm{B}$ is perpendicular to $\bm{g}$, then the kinetic
pressure gradient $dp_\perp^\mathrm{kin}/dz$ 
acts in parallel with
the Lorentz force density $BdM/dz$. Note that in
our case $dM/dz \neq 0$, simply because $d\rho/dz \neq 0$
($\rho$ depends on $z$) in the gravity field.
Since $B$ and $T$ are constant,
\beq
 \frac{dM}{dz}=\frac{\partial M(\rho,T,B)}{\partial\rho}\,
    \frac{d\rho}{dz}
    = - \frac{\partial^2\Omega(\rho,T,B)
       }{
          \partial\rho\, \partial B}\,
    \frac{d\rho}{dz} .
\label{dMdz}
\eeq
Then the equilibrium condition takes the same standard form
\[
 \rho g = \frac{dp_\perp^\mathrm{kin}}{dz} +B\frac{dM}{dz}
=\frac{d}{dz}(-\Omega-MB)+B\frac{dM}{dz}=-\frac{d\Omega}{dz}.
\]

Thus, the gradient
$d\rho/dz=-(\partial\Omega/\partial\rho)^{-1}\rho g$
does not depend on
$\bm{B}$-field direction, which means that the hydrostatic
equilibrium is determined by the isotropic thermodynamic pressure
$P$, in accordance with the results of Ref.~\cite{bh82}.

Since the forces created by bound currents are small in the
majority of applications, the equations of
magnetohydrodynamics (MHD) are commonly derived neglecting
the magnetization. However, the magnetization term is easily
recovered by substituting  the general expression $\bm{j} =
\bm{j}_f + \bm{j}_b$ into the microscopic Lorentz force
density $\bm{j}\times\bm{B}/c$ that is included in the
derivation of MHD equations from the first principles (e.g.,
\cite{LaLi-ECM}, Chap.~VIII). Moreover, thermodynamics of
magnetized media is well studied in the theory of magnetics
(e.g., \cite{LaLi-ECM}, Chap.~IV). Of course, everyone is
free to use anisotropic kinetic pressure in MHD equations
and add the magnetization force density explicitly. However,
it seems more natural to follow the traditional approach and
use the isotropic thermodynamic pressure that automatically
includes the contribution of the magnetization.

\begin{acknowledgments}
We thank Andrey Chugunov for drawing our attention to
Ref.~\cite{ferreretal10} and acknowledge partial support
from the RFBR Grant 11-02-00253-a and the Russian Leading
Scientific Schools program (Grant NSh-3769.2010.2). DY acknowledges
also support from the RFBR Grant 11-02-12082-ofi-m-2011 and from
Ministry of Science and Education of Russia (contract
11.G34.31.0001).
\end{acknowledgments}


\begin{thebibliography}{222}

\bibitem{ferreretal10}
E.~J.~Ferrer, V.~de la Incera, J.~P.~Keith, I.~Portillo, and
P.~L.~Springsteen, Phys.\ Rev.\ C {\bf 82}, 065802 (2010).

\bibitem{ChandraFermi}
S.~Chandrasekhar and E.~Fermi,
Astrophys.~J. \textbf{118}, 116 (1953);
erratum: \textit{ibid.}, \textbf{122}, 208 (1955).

\bibitem{LaiShapiro}
D.~Lai and S.~L. Shapiro,
Astrophys.~J. \textbf{383}, 745 (1991).

\bibitem{Bocquet}
M.~Bocquet, S.~Bonazzola, E.~Gourgoulhon, and J.~Novak,
Astron.\ Astrophys. \textbf{301}, 757 (1995).

\bibitem{KK08}
K.~Kiuchi and K.~Kotake,
Mon.\ Not.\ R.\ astron.\ Soc. \textbf{385}, 1327 (2008).

\bibitem{cc68}
V.~Canuto and H.-Y.~Chiu, Phys.\ Rev.\ \textbf{173}, 1210
(1968); \textbf{173}, 1220 (1968).

\bibitem{bh82}
R.~D.~Blandford and L.~Hernquist, J.~Phys.~C \textbf{15},
6233 (1982).

\bibitem{Griffith}
D.~J. Griffith,
\textit{Introduction to Electrodynamics}, 3rd ed.
(Prentice-Hall, London, 1999), Chap.~6.

\bibitem{LaLi-ECM}
L.~D. Landau, E.~M. Lifshitz, and L.~P. Pitaevski\u{\i},
\textit{Electrodynamics of Continuous Media}, 2nd ed.
(Butterworth-Heinemann, Oxford, 1984).

\end{thebibliography}
\end{document}